# The unexpected binding and superconductivity in $SbH_4$ at high pressure

*Yanbin Ma, Defang Duan, Da Li, Yunxian Liu, Fubo Tian, Xiaoli Huang, Zhonglong Zhao, Hongyu Yu, Bingbing Liu, Tian Cui\**

*State Key Laboratory of Superhard Materials, College of Physics, Jilin University, Changchun 130012, People's Republic of China*

**Abstract**

The semimetal antimony (Sb) element doped into hydrogen has been performed theoretically to explored high-pressure crystal structures and superconductivity of antimony hydrides. The unexpected stoichiometry $SbH_4$ with $P6_3/mmc$ symmetry is found to have most negative enthalpy and embody the coexistence of covalent and ionic bonds. It is a metallic phase and stable in the pressure range of 127-300 GPa. Furthermore, a superconducting critical temperature ($T_c$) of 106 K is obtained at 150 GPa by employing the Allen-Dynes modified McMillan equation. In addition, an extrusive distinguishing feature is the presence of soft phonon modes which is primary contribution to the strength of electron-phonon coupling.

The elusive holy grail of high-pressure physics that hydrogen can realize metallization under high pressure had been put forward eight decades ago [1]. Since then, many scientists have been devoted to studying the metallization of hydrogen and the steps will never stop. However, even pressure up to 360 GPa [2], the expected topic has been not observed. By doping the heavier elements into hydrogen, hydrogen-rich compounds are considered to metalize at lower pressures than pure hydrogen because of the "chemical precompression" proposed by N. W. Aschroft in 2004 [3]. Moreover, they hold out the hope that the hydrogen-rich compounds are the potential high-temperature superconductors. For instance, *Duan et al.* have proposed that the $H_3S$ [4, 5] with *Im-3m* symmetry produces high $T_c$ values of 191 ~ 204 K at 200 GPa which have been proved through the high-pressure experiment [6]. Other hydrides, alkali metal hydride $KH_6$ [7], the sodalite-like cage structure $CaH_6$ [8] and $YH_6$ [9], the group IIIA metal hydride $GaH_3$ [10], the group IVA hydrides $Si_2H_6$ [11], $GeH_4$ [12], $SnH_4$ [13] were predicted to possess high $T_c$ under high pressures. It may be a potential rule that large hydrogen fraction may be an essential condition to produce the high $T_c$ for hydrogen-rich compounds.

As the group VA element, the nitrogen hydrides have been studied widely, but they have difficulty in achieving metallization under high pressure [14, 15]. Unluckily, the hydrides of phosphorus, arsenic and antimony are seldom to be studied under high pressure. Antimony (Sb) element has a greater mass and more electrons than Sn which may offer more electrons to Fermi level of hydrides and promote superconductivity after metallization. In addition, given the electronegativity, the numerical value of Sb (2.05) is greater than Sn (1.96) and close to H (2.20), which might lead the Sb to combine with H and form stoichiometry $SbH_4$. Particularly, the alike electronegativity metal tellurium (2.10) combining with hydrogen was predicted to form $TeH_4$ [16].

In order to search potential high-temperature superconductivity in H-rich regime of Sb-H compounds, we explored their crystal structure at high pressure by considering six components $SbH_n$ (n=1~6). Strikingly, $SbH_4$ with $P6_3/mmc$ has the most negative enthalpy of formation from 127 to 300 GPa. It contains quasi-molecular $H_2$ units and the coexistence of covalent and ionic bonds. Further studies reveal that it is a high-temperature superconductor with $T_c$ of 106 K at 150 GPa.

Structural prediction was performed with the USPEX code [17,18,19], which is used widely to predicted crystalline structures [20,21,22]. Besides, we hunt for high-pressure crystal structure of Sb by applying CALYPSO code [23, 24], which is also very successful in predicting the elements [25, 26] and compounds [27]. The structural optimization was executed with the Perdew–Burke–Ernzerhof of generalized gradient approximation [28] and the all-electron projector-augmented wave method [29], implemented in the Vienna ab initio simulation package VASP code [30]. The most optimal structures were further re-optimized at higher accurate level: a plane-wave basis set cutoff of 800 eV and a integrated Brillouin zone sampling grid spacing of $2\pi \times 0.03$ Å$^{-1}$. The lattice dynamics and superconducting properties of $SbH_4$ were calculated by means of density functional perturbation theory [31], as implemented in the QUANTUMESPRESSO code [32]. The norm-conserving pseudopotentials for H and Sb were used. To ensure the convergence tests, we chose an 80 Ry kinetic energy cutoff and a $16 \times 16 \times 16$ Monkhorst-Pack (MP) [33] $k$-point sampling mesh. A $4 \times 4 \times 4$ $q$-mesh in the first BZ was used to insert into the force constants to evaluate the phonon band structure. More computational details are supplied in Supplementary materials (Sm) [34].

The formation enthalpy of $SbH_n$ (n=1~6) relative to dissociation into the products of Sb and $H_2$ at 0 K and 100, 150, 200, 250, 300 GPa are calculated and summarized in Fig. 1. It can be seen that no component is stable against elemental dissociation below 100 GPa, conforming the fact of inexistence of Sb hydrides at low pressures. Further compression to 127 GPa, the $P6_3/mmc$ of $SbH_4$ becomes the most stable structure (see Fig. S1 and S2 in Sm). With increasing pressure to 150 GPa, all the stoichiometries become energetically favored. The SbH compound stabilizes on convex hull at 200 GPa with $T_c$=10.5 K (5.6 K at 300 GPa) when $\mu^*$ is selected 0.10. The $SbH_3$ compound is stable on tie-line at 300 GPa. The SbH and $SbH_3$ are dynamic stability in their stable pressure ranges, as depicted in Fig. S3, S5 and S6 of Sm [34].

In many hydrides, the MH$_x$ ($x$ is equal to the number of valence electrons), such as LiH [35], NaH [36], KH [37], MgH$_2$ [38], CaH$_2$ [8], AlH$_3$ [39], GaH$_3$ [10] is usually the most stable stoichiometry even at high pressures. However, in our work the SbH$_4$ has the most negative enthalpy in all researched antimony hydrides. Considering electronic configuration of Sb atom, the most stable SbH$_4$ is against the conventional SbH$_3$ or SbH$_5$ meaning that high pressure may alter the valence electronic state. In other words, the SbH$_4$ may be more easily synthesized in experiment by compressing Sb+H$_2$.

In the $P6_3/mmc$ of SbH$_4$, it is very highlighted that the "H$_2$" units arranging along $C$ axis compose a hexagonal frame where two Sb atoms and four H atoms occupy on 2$c$ and 4$f$ sites, respectively, as depicted in Fig. 2(a), S3 (c). The H$_2$-contained SbH$_4$ is similar with IVA hydrides (GeH$_4$ [12] and SnH$_4$ [40]), but there is a deviation of bonding form that will be discussed below. Detailed crystallographic information of SbH$_4$ at 150 GPa are listed in Table SI and Table SII of Sm. It is shown that the distance between H atoms in the H$_2$ units decreases from 0.85 (150 GPa) to 0.832 Å (300 GPa) which is longer than pure hydrogen at same pressure.

To understand the electronic properties, we calculate the electronic band structure and density of states (DOS). As can be seen in Fig 3 (a), the overlap of valence bands and conduction bands at 150 GPa implies that the $P6_3/mmc$ of SbH$_4$ is a good metal. The DOS displays that near the Fermi level the orbits of Sb and H atoms hybridize strongly, which hints a crucial signal that the $P6_3/mmc$ SbH$_4$ may have a large degree of Fermi surface filling in the first Brillouin zone, as depicted in illustration of Fig.3 (b). I. It is conducive to enhancing the electron-phonon coupling (EPC). Besides, the $P6_3/mmc$ of SbH$_4$ have a larger DOS (4.64×10$^{-2}$ states/eV/Å$^3$) than SnH$_4$ [40] with $P6_3/mmc$ symmetry (2.18×10$^{-2}$ states/eV/Å$^3$) at 200 GPa. This larger DOS may account for a higher $T_c$.

It is very clear that each Sb atom loses 1.282 $e$ by calculating the Bader charges (Table SV of Sm) at 150 GPa. However, where are the losing electrons and what will they do? It is not adequate to analyze only from the losing charges. Therefore, the electron localization function (ELF) was calculated to explore bonding form. As shown in Fig. 2 (c) and (d), the value of ELF between the nearest H atoms arranging along C axis is close to 0.95, suggesting the existence of $H_2$ molecular units. In addition, the nearest Sb and H atoms form startling weak polar covalent bonds when the isosurface value of 0.7 is selected and fades in 0.75. When the $SbH_4$ is further squeezed up to the pressure 300 GPa, the losing charges reach 1.492$e$. Although each Sb atom loses more charges (Table SV of Sm) and the nearest separations between Sb and H decrease with the increasing pressure, which does not make the covalent bonding trend to become more obvious, instead of more weak (depicted in Fig. S4 of Sm). The ELF interprets that it is not simple for Sb atoms to donate electrons to H atoms and boost the formation of ionic H. That is different from other hydrogen-rich compounds, for example $GaH_3$ [10], $Si_2H_6$ [11] and $TeH_4$ [16], where the heavier atoms provides their electrons to H atoms and form the ionic H. Even more intriguing discrepancy is that the $SbH_4$ is diverse from group IVA hydrides (i.e., $SiH_4$, $GeH_4$, and $SnH_4$), where some hydrogen atoms are stronger covalently bonded to M (M=Si, Ge, or Sn) atoms to form M-H bonds. That is to say, $SbH_4$ is the hydride with coexistence of covalent and ionic bonds. The pressure plays a crucial role on the evolution of bonding form. With the increasing pressure, the covalent bonds weaken, while the ionic bonds strengthen. To the best of our knowledge, Sb is a semimetal situated within boundary of the group VA metal and non metal, which may lead to Sb possess the nature between the metal and non metal. Moreover, the electronegativity of Sb is 2.05 slightly smaller than H (2.20), indicating that H atoms may have a stronger ability to attract electrons than Sb and results in the unusual bonding form.

In order to explore the superconductivity of the $SbH_4$, the phonon frequency logarithmic average ($\omega_{\log}$), DOS at Fermi level $N(E_f)$ and EPC parameter ($\lambda$), at different pressure are calculated and listed in Table SIII [34]. According to our calculations, $\omega_{\log}$ is 1118.60 K and $\lambda$ reaches a high value 1.26 at 150 GPa, resulting that the $T_c$ achieves 95-106 K evaluated with the Allen-Dynes modified McMillan equation [41] $T_C = \frac{\omega_{\log}}{1.2} \exp[-\frac{1.04(1+\lambda)}{\lambda - \mu^*(1+0.62\lambda)}]$ with $\mu^*$ of 0.13-0.10. With the increasing pressure, $T_c$ decreases slowly (106 K at 150 GPa and 82 K at 300 GPa for $\mu^*$ =0.10) at a descendant gradient ($dT_c/dP$) of -0.16 K/GPa. The $T_c$ of $P6_3/mmc$-SbH$_4$ has only a weak dependence on pressure.

As a function of the frequency for the $P6_3/mmc$ at 150 GPa, the related phonon density of states, Eliashberg spectral function $\alpha^2F(\omega)$ and integrated $\lambda$ are presented in Fig. 4. No imaginary frequency in the phonon dispersion curves suggests that SbH$_4$ is dynamically stable. It is shown that Sb atomic vibrations in the frequency region below 10 THz contribute approximately 7% to total $\lambda$, while intermediate-frequency modes (between 16 and 52 THz) contribute 92% of $\lambda$ and contribution of the intramolecular (H$_2$ units) vibration modes is only 1% or less. It is very obvious that the superconductivity of SbH$_4$ with $P6_3/mmc$ symmetry is mainly attributed by intermediate-frequency. This superconductive mechanism is akin to SnH$_4$ [13, 40], GaH$_3$ [10], TeH$_4$ [16] and PoH$_4$ [42].

The remarkable high contribution of 92% drives us to explore the superconducting mechanism of SbH$_4$ from specific vibration modes. As depicted in Fig. 4 (a), we calculated electron-phonon coupling strength projected on each vibration mode and the radius sizes of orange solid balls represent the strength. Comparing Fig. 3(a) and Fig. 4(a), the region of electronic bands crossing Fermi surface along Brillouin zone Γ-M, Γ-K and M-L corresponds to soft phonons. It leads to the strong EPC to one or several particular branches in the restricted regions of momentum space determined by the FS topology (Fig. 3(b)). From the calculated EPC strength projected on each vibration mode, we found that the $\lambda$ is contributed mainly by low frequency of optical soft phonon modes, along the Brillouin zone Γ-M, Γ-K and M-L. The 3D Fermi surfaces inserted in Fig. 3(b) qualitatively expounds the Fermi-surface nesting. We found that the nesting might happen along Γ-M, Γ-K and M-L corresponding to the soft phonons and high value $\lambda$. Particularly, it can also be viewed that all the phonon branches harden when the pressure reaches 300 GPa, which may cause the $T_c$ to decrease.

In conclusion, a hexagonal structure SbH$_4$ with the lowest enthalpy of formation is revealed for the first time. It takes on the characteristics of metal as coexistence of covalent and ionic bonds and is stable above 127 GPa. Exhilaratingly, a high $T_c$ of 95-106 K at 150 GPa was predicted. The high $T_c$ is closely relevant to the soft modes of low-frequency optical branches. The predicted $SbH_4$ sustains the standpoint that hydrogen-rich materials provide an effective way to achieve metallization and high temperature superconductor are hoped to come true at accessible experimental pressure. The present work will also encourage more and more scientists to study superconductivity of the group VA hydrogen-rich compounds.

This work was supported by the National Basic Research Program of China (No. 2011CB808200), Program for Changjiang Scholars and Innovative Research Team in University (No. IRT1132), National Natural Science Foundation of China (Nos. 51032001, 11204100, 11074090, 10979001, 51025206, 11104102 and 11404134), National Found for Fostering Talents of basic Science (No. J1103202), China Postdoctoral Science Foundation (2012M511326, 2013T60314 and 2014M561279). Parts of calculations were performed in the High Performance Computing Center (HPCC) of Jilin University.

**Captions**

Fig.1. (Color online) The enthalpies of formation (with respect to Sb and $H_2$) of $SbH_n$ (n=1~6). Symbols for the full filled denote the structures are on the tie-lines, but for half filled represents that are not on tie-lines.

Fig.2 (Color online) The calculated ELF of *P6$_3$/mmc* $SbH_4$ at 150 GPa (a) isosurface value of 0.70, (b) isosurface value of 0.75, (c) for the (110) plane, (d) for the (1-10) plane. Large purple spheres represent Sb and small green spheres denote H atoms, respectively.

Fig.3. (Color online) (a) The electronic band structure (left) and density of states (right) of *P6$_3$/mmc*-$SbH_4$ at 150 GPa. The dotted lines at zero indicate the Fermi level. (b) Fermi surface topology of *P6$_3$/mmc*-$SbH_4$ at 150 GPa, (I) the 3D view of the Fermi surface including all cutting bands and (II)-(VII) the Fermi surface crossing the Fermi energy. The Fermi surface is calculated with a $24 \times 24 \times 12$ *k* mesh.

Fig.4. (Color online) (a) The calculated phonon dispersion curves for *P6$_3$/mmc*-$SbH_4$ at 150 GPa with the radius sizes of orange solid balls representing the strength of EPC. (b) The phonon DOS projected on H and Sb atoms. (c) The Eliashberg phonon spectral function $\alpha^2F(\omega)$ and the partial electron-phonon integral $\lambda(\omega)$.

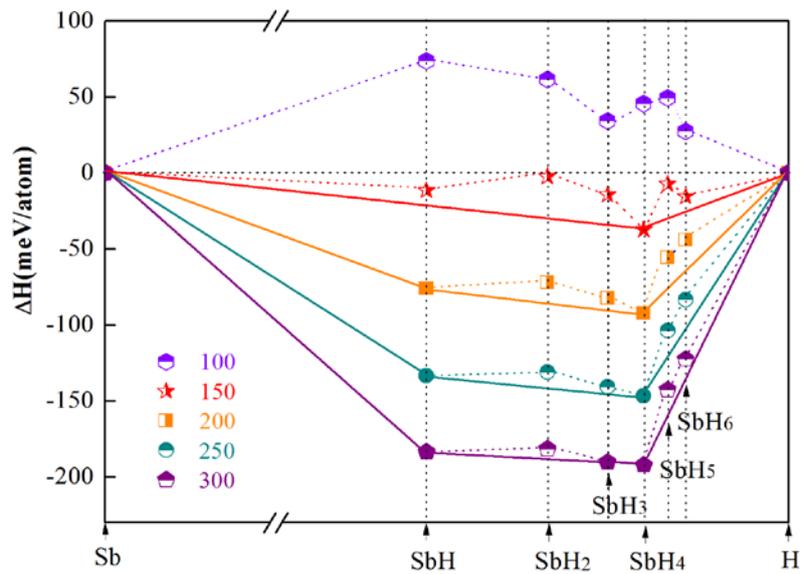

**FIG. 1**

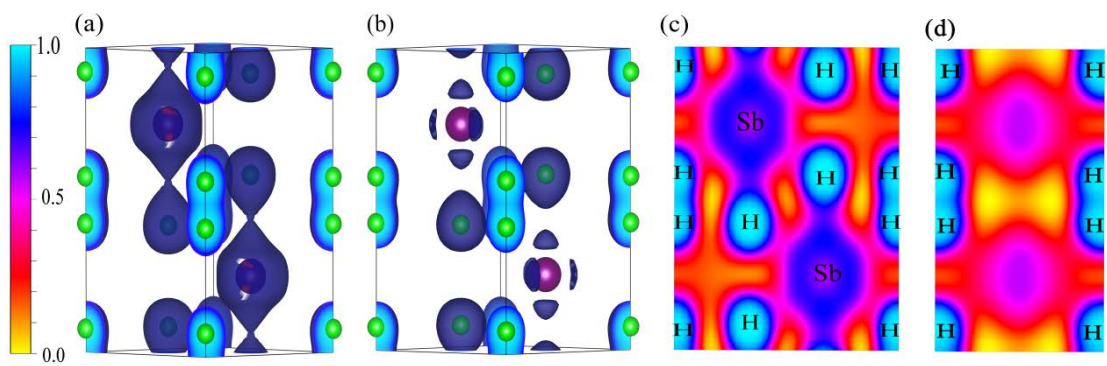

FIG. 2

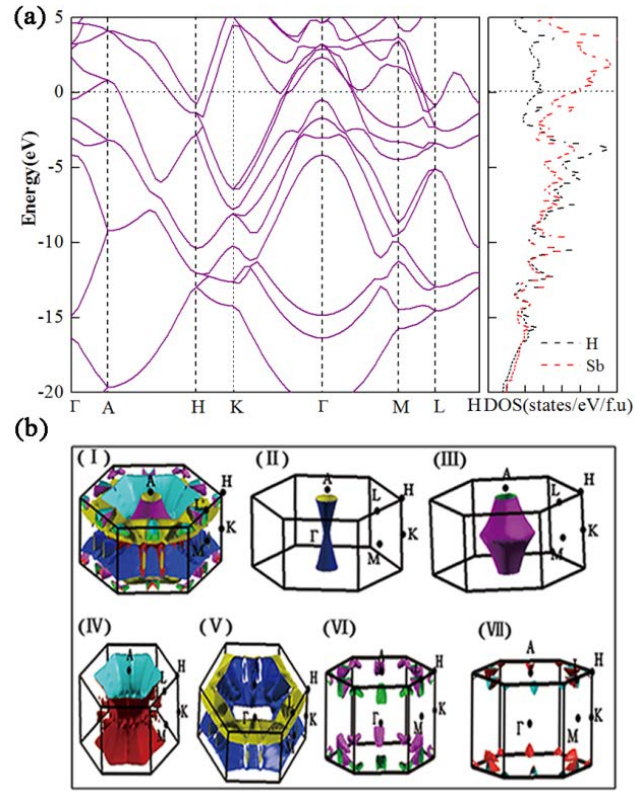

**FIG. 3**

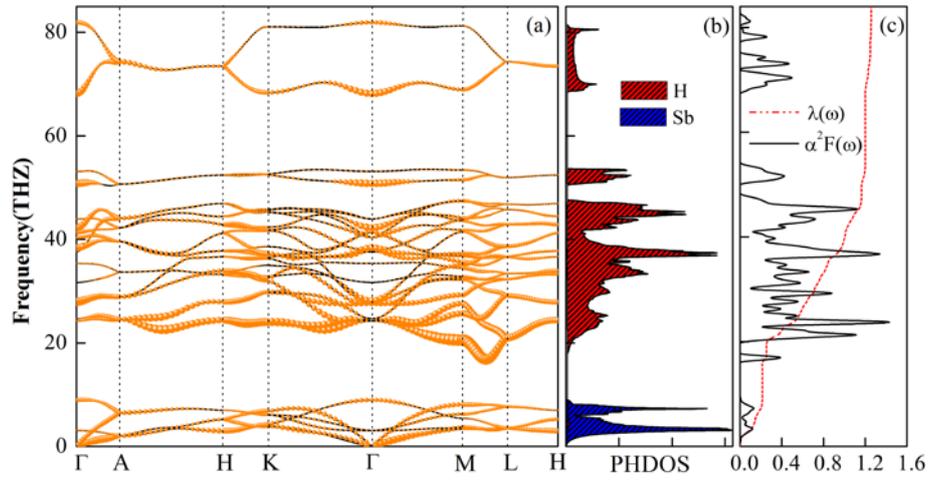

**FIG. 4**

# Supplementary Material

# The unexpected binding and superconductivity in SbH$_4$ at high pressure

*Yanbin Ma, Defang Duan, Da Li, Yunxian Liu, Fubo Tian, Xiaoli Huang, Zhonglong Zhao, Hongyu Yu, Bingbing Liu, Tian Cui\**

*State Key Laboratory of Superhard Materials, College of Physics, Jilin University, Changchun 130012, People's Republic of China*

# I Computational details

The high-pressure crystal structures of antimony hydrides were explored by using the evolutionary algorithm and *ab initio* total-energy calculations, as implemented in the USPEX code [1,2,3], considering simulation sizes ranging from one to four antimony hydrides formula units (f.u.) at 50, 100, 150, 200, 250 and 300 GPa. The high-pressure structures of Sb were hunted by applying CALYPSO code [4, 5], considering simulation sizes ranging from one to eight formula units (f.u.) at 50, 100, 150, 200, 250 and 300 GPa. We found that a plane-wave basis set cutoff of 500 eV and a integrated Brillouin zone (BZ) sampling grid spacing of $2\pi \times 0.05^{-1}$ Å are enough to predict the high-pressure crystal structures of $SbH_n$ (n=1~6). The optimal energetically structures are re-optimized to attain the convergence criterion that all forces on atoms are converged to less than 0.005 eV/Å. H (1s) and Sb ($5s^2p^3$) were treated as valence electrons. The phonon of SbH was calculated by QUANTUMESPRESSO code [6] with an 80 Ry kinetic energy cutoff, a $12 \times 24 \times 18$ Monkhorst-Pack (MP) and a $2 \times 4 \times 3$ *q*-mesh in the first BZ. The phonon dispersion curve of $SbH_3$ was calculated by Cambridge Sequential Total Energy Package (CASTEP) with a $4 \times 4 \times 4$ super cell [7].

# II Calculation of superconducting transition temperature

The linewidth of a phonon mode engendering from electron–phonon interaction is expressed by

$$\gamma_{qv} = 2\pi\omega_{qv} \sum_{mn}\sum_{k} \left|g_{k+q,k}^{qv,mn}\right|^2 \delta(\varepsilon_{k+q,m} - \varepsilon_F)\delta(\varepsilon_{k,n} - \varepsilon_F) \qquad (1)$$

Where $\left|g_{k+q,k}^{qv,mn}\right|^2$ is square of the electron–phonon matrix element and the sum is over the Brillouin zone (BZ), and $\varepsilon_k$ are the energies of bands determined with respect to the Fermi level at point $k$. So, the electron-phonon coupling (EPC) spectral function $\alpha^2 F(\omega)$ can be given by relational expression of the phonon linewidth $\gamma_{qv}$, as a result of electron-phonon scattering,

$$\alpha^2 F(\omega) = \frac{1}{2\pi N(\varepsilon_F)} \sum_{qv} \frac{\gamma_{qv}}{\omega_{qv}} \delta(\omega - \omega_{qv}) \qquad (2)$$

Where $N(\varepsilon_F)$ is the electronic density of states (DOS) at the Fermi surface. A Gaussian of 0.03 was tested and appropriate to displace the $\delta$ function. The electron-phonon coupling (EPC) parameter $\lambda$ is expressed as

$$\lambda = 2\int_0^\infty \frac{\alpha^2 F(\omega)}{\omega} d\omega \qquad (3)$$

The superconducting transition temperature $T_C$ has been estimated with the Allen-Dynes modified McMillan equation as [7]

$$T_C = \frac{\omega_{\log}}{1.2} \exp\left[-\frac{1.04(1+\lambda)}{\lambda - \mu^*(1+0.62\lambda)}\right] \qquad (4)$$

Where $\omega_{\log}$ is the logarithmic average frequency that can be directly obtained by calculating the phonon dispersion curves, and $\mu^*$ is the Coulomb pseudopotential where $\mu^*=0.10\sim0.13$ is suitable value for hydrogen-rich compounds.

# III Figures

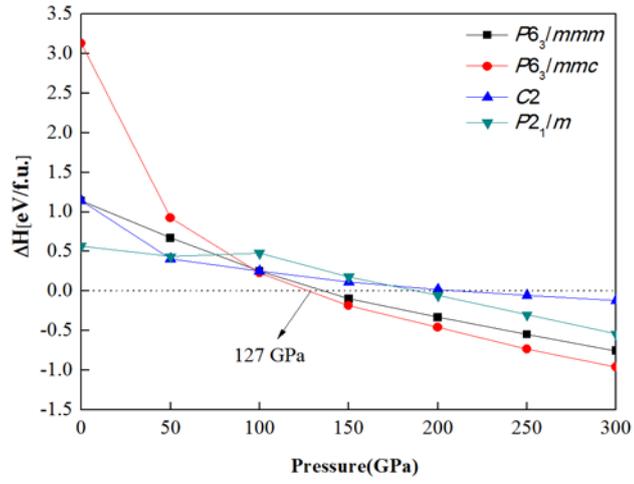

**FIG.S1** (color online) The calculated enthalpies of formation as a function of pressure for predicted SbH$_4$ structures. △H=H (SbH$_4$) – H (Sb) -2H (H$_2$).

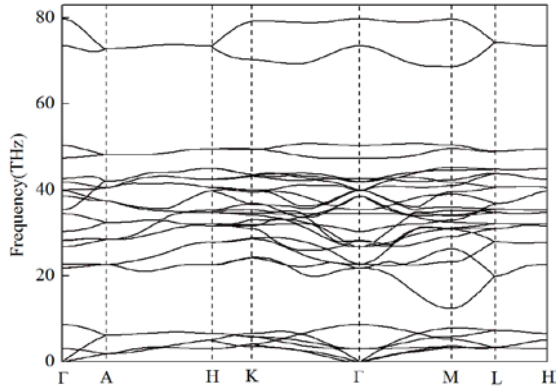

**Fig.S2.** (Color online) The calculated phonon band structure for $P6_3/mmc$-SbH$_4$ at 127 GPa.

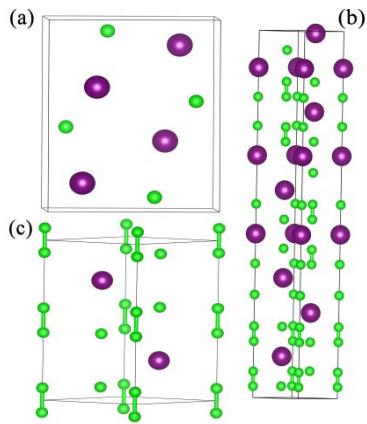

**Fig.S3** (Color online) (a) The SbH with *Pnma* symmetry at 200 GPa. (b) The *R3m* SbH$_3$ at 300 GPa. (c) The $P6_3/mmc$ of SbH$_4$ at 150 GPa. Large purple spheres represent Sb and small green spheres denote H atoms, respectively.

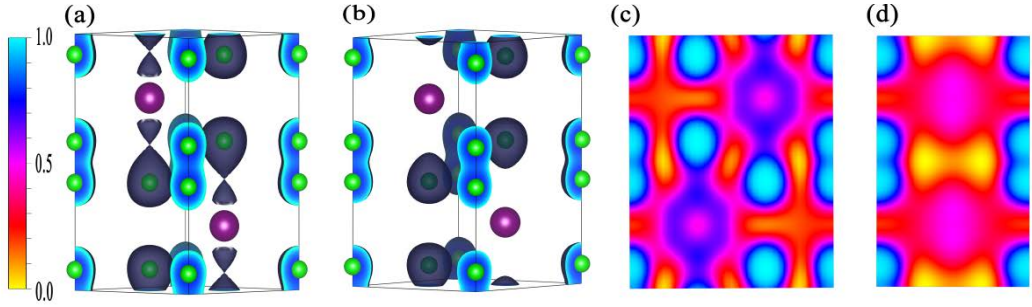

**Fig.S4** The calculated ELF of $P6_3/mmc$-SbH$_4$ at 300 GPa (a) with isosurface value of 0.70, (b) with isosurface value of 0.75, (c) for the (110) plane. (d) for the (1-10) plane. Large purple spheres represent Sb and small green spheres denote H atoms, respectively.

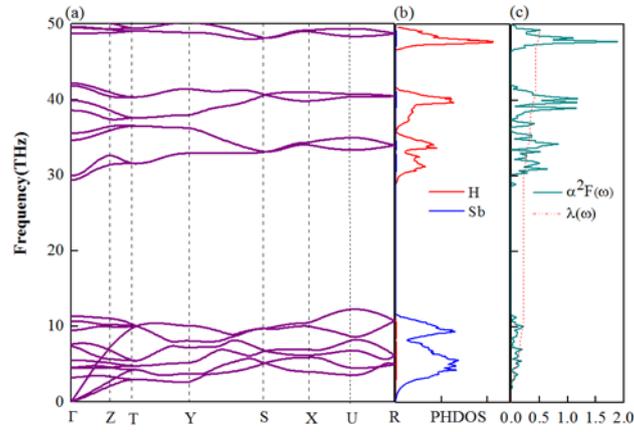

**Fig.S5.** (Color online) (a) The calculated phonon band structure for *Pnma* SbH at 200 GPa. (b) The phonon DOS projected on Sb and H atoms. (c) The Eliashberg phonon spectral function, $\alpha^2F(\omega)$ and the partial electron-phonon integral, $\lambda(\omega)$.

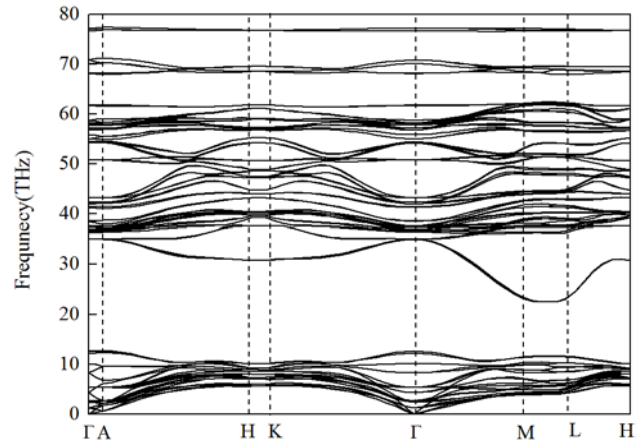

**Fig.S6.** (Color online) Calculated phonon band structure for $R3m$ SbH$_3$ at 300 GPa.

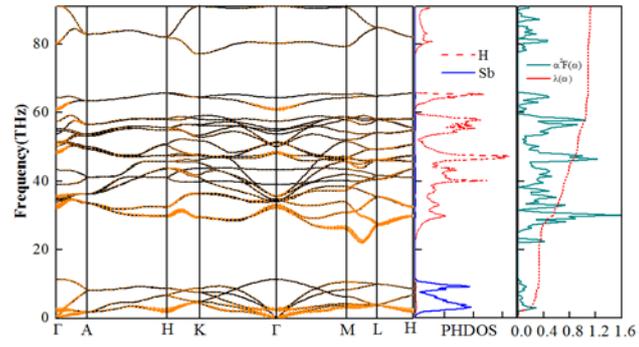

**Fig.S7.** (Color online) (a) Calculated phonon band structure for $P6_3/mmc$ SbH$_4$ at 300 GPa. Orange solid circles indicate the electron-phonon coupling with the radius proportional to their respective strength. (b) The phonon DOS projected on Sb and H atoms. (c) The Eliashberg phonon spectral function, $\alpha^2F(\omega)$, and the partial electron-phonon integral, $\lambda(\omega)$.

# IV TABLE

**Table SI**. The lattice parameters and atomic positions of Sb at 0, 50, 100 GPa (predicted by Calypso), SbH at 200 GPa, SbH$_3$ at 300 GPa and SbH$_4$ at 150 GPa.

| Structure #Space #Pressure | Parameters (Å, deg) | Atom | x | y | z |
|---|---|---|---|---|---|
| Sb *R-3m* (0 GPa) | a=4.3991 b=4.3991 c=11.4680 γ=120 | Sb1 | 0.00000 | 0.00000 | 0.23170 |
| Sb *Im-3m* (50 GPa) | a=b=c=3.36 | Sb1 | 0.00000 | 0.00000 | 0.00000 |
| Sb *Fmmm* (100 GPa) | a=3.1764 b=4.5402 c=4.5405 | Sb1 | 0.00000 | 0.50000 | 0.500000 |
|  |  | Sb2 | 0.00000 | 0.00000 | 0.000000 |
|  |  | Sb3 | 0.50000 | 0.50000 | 0.000000 |
|  |  | Sb4 | 0.50000 | 0.00000 | 0.500000 |
| SbH Pnma (200 GPa) | a=5.3172 b=2.6941 c=4.2262 | H1 | 0.42776 | -0.75000 | 0.12996 |
|  |  | Sb1 | 0.35493 | 0.25000 | 0.70880 |
| SbH$_3$ *R3m* (300 GPa) | a=5.3172 b=2.6941 c=4.2262 α=β=γ=90 | H1 | 1.03120 | 1.03120 | 0.03120 |
|  |  | H2 | 1.15365 | 1.15365 | 0.15365 |
|  |  | H3 | 1.28069 | 1.28069 | 0.28069 |
|  |  | H4 | 1.19470 | 1.09470 | 0.19470 |
|  |  | H5 | 1.07152 | 1.07152 | 0.07152 |
|  |  | H6 | 0.74054 | 0.74054 | 0.74054 |
|  |  | H7 | 0.81937 | 0.81937 | 0.81937 |
|  |  | H8 | 1.52766 | 0.52766 | 0.52766 |
|  |  | H9 | 1.36535 | 0.36535 | 0.36535 |
|  |  | Sb1 | 0.89921 | 0.89921 | 0.89921 |
|  |  | Sb2 | 1.66025 | 0.66025 | 0.66025 |
|  |  | Sb3 | 1.44651 | 0.44651 | 0.44651 |
| SbH$_4$ *P6$_3$/mmc* (150 GPa) | a= 2.9981 b= 2.9981 c= 5.5040 γ= 120 | H1 | 0.66667 | 0.33333 | 0.41682 |
|  |  | H2 | 1.00000 | 0.00000 | 0.42275 |
|  |  | Sb1 | 0.33333 | 0.66667 | 0.25000 |

**Table SII.** The nearest distances of Sb-Sb, Sb-H, Sb-H$_2$, H$_2$-H$_2$ and H-H in the "H$_2$" units of SbH$_4$ with $P6_3/mmc$ symmetry at 150, 200, 250, and 300GPa.

| Pressure (GPa) | Sb-Sb (Å) | Sb-H (Å) | Sb-H$_2$ (Å) | H$_2$-H$_2$ (Å) | H-H (Å) |
|---|---|---|---|---|---|
| 150 | 2.998 | 1.834 | 1.975 | 1.902 | 0.850 |
| 200 | 2.914 | 1.189 | 1.918 | 1.841 | 0.846 |
| 250 | 2.842 | 1.755 | 1.873 | 1.806 | 0.840 |
| 300 | 2.786 | 1.725 | 1.837 | 1.773 | 0.832 |

**Table SIII.** The phonon frequency logarithmic average ($\omega_{\log}$), EPC parameter ($\lambda$), the electronic DOS at Fermi level $N(E_f)$, and critical temperature $T_c$ ($\mu^* = 0.1$ and 0.13) for SbH$_4$ at 150, 200, 250 and 300 GPa.

| P (GPa) | Lambda ($\lambda$) | $\omega_{\log}$(K) | N($E_f$) states/Spin/Ry/cell | Tc (K) with $\mu^*$ = 0.1, 0.13 | |
|---|---|---|---|---|---|
| 150 | 1.26 | 1118.60 | 8.11 | 106 | 95 |
| 200 | 1.17 | 1134.02 | 7.74 | 98 | 87 |
| 250 | 1.13 | 1096.54 | 7.28 | 90 | 79 |
| 300 | 1.12 | 998.36 | 6.70 | 82 | 72 |

**Table SIV.** The phonon frequency logarithmic average ($\omega_{\log}$), EPC parameter ($\lambda$), the electronic DOS at Fermi level $N(E_f)$, and critical temperature $Tc$ ($\mu^* = 0.1$ and 0.13) for SbH at 200 and 300 GPa.

| P (GPa) | Lambda($\lambda$) | $\omega_{\log}$ (K) | N($E_f$) states/Spin/Ry/cell | Tc (K) with $\mu^*$ = 0.1, 0.13 | |
|---|---|---|---|---|---|
| 200 | 0.50 | 853.93 | 9.92 | 10.5 | 6.3 |
| 300 | 0.43 | 966.71 | 9.28 | 5.6 | 2.7 |

**Table SV**. Calculated Bader charges of H and Sb atoms in SbH$_4$ (*P6$_3$/mmc*) at 150 and 300 GPa.

| Pressure(GPa) | Atom | Charge | σ(e) |
|---|---|---|---|
| 150 | H1 | 1.4291 | -0.4291 |
|  | H2 | 1.4285 | -0.4285 |
|  | H3 | 1.4291 | -0.4291 |
|  | H4 | 1.4287 | -0.4287 |
|  | H5 | 1.2401 | -0.2401 |
|  | H6 | 1.1846 | -0.1846 |
|  | H7 | 1.2401 | -0.2401 |
|  | H8 | 1.1846 | -0.1846 |
|  | Sb1 | 3.7176 | 1.2824 |
|  | Sb2 | 3.7176 | 1.2824 |
| 300 | H1 | 1.4790 | -0.4790 |
|  | H2 | 1.4780 | -0.4780 |
|  | H3 | 1.4780 | -0.4780 |
|  | H4 | 1.4790 | -0.4790 |
|  | H5 | 1.2949 | -0.2949 |
|  | H6 | 1.2401 | -0.2401 |
|  | H7 | 1.2948 | -0.2948 |
|  | H8 | 1.2401 | -0.2401 |
|  | Sb1 | 3.5080 | 1.4920 |
|  | Sb2 | 3.5081 | 1.4920 |